# Deuterium retention and thermal conductivity in ion-beam displacement-damaged tungsten

G.R. Tynan *[1,2], R.P. Doerner[1], J. Barton[1,2], R. Chen[1,2], S. Cui[2], and M. Simmonds[1,3]
[1]Center for Energy Research
[2]Department of Mechanical and Aerospace Engineering, Jacobs School of Engineering
[3]Department of Physics
UC San Diego, La Jolla CA USA

Y. Wang, J.S. Weaver, N. Mara, S. Pathak[4]
Ion Beam Materials Laboratory
Center for Integrated Nanotechnologies
Los Alamos National Laboratory, Los Alamos NM USA

[4]Current affiliation:
Dept. of Chemical and Materials Engineering
University of Nevada, Reno
Reno, NV 89557

## ABSTRACT

Retention of plasma-implanted D is studied in W targets damaged by a Cu ion beam at up to 0.2 dpa with sample temperatures between 300 K and 1200 K. At a D plasma ion fluence of $10^{24}/m^2$ on samples damaged to 0.2 dpa at 300 K, the retained D retention inventory is $4.6 \times 10^{20}$ D/m$^2$, about ~5.5 times higher than in undamaged samples. The retained inventory drops to $9 \times 10^{19}$ D/m$^2$ for samples damaged to 0.2 dpa at 1000 K, consistent with onset of vacancy annealing at a rate sufficient to overcome the elevated rate of ion beam damage; at a damage temperature of 1200 K retention is nearly equal to values seen in undamaged materials. A nano-scale technique provides thermal conductivity measurements from the Cu-ion beam displacement damaged region. We find the thermal conductivity of W damaged to 0.2 dpa at room temperature drops from the un-irradiated value of 182±3.3 W/m·K to 53±8 W/m·K.

* corresponding author. 460 EBU II MC 0417, UC San Diego, 9500 Gilman Drive, La Jolla CA 92093

# 1.0 Introduction:

Fusion engineering test reactor concepts that demonstrate steady-state reactor-relevant operations with a closed tritium fuel cycle and begin to produce useful energy are of interest[1-3], and are expected to have material surfaces that must withstand intense heat fluxes reaching 10 $MW/m^2$ or more at the divertor target with particle fluxes of $10^{19}$ - $10^{20}$ $/m^2$-sec at the first wall and $10^{23}$-$10^{24}$ $/m^2$-sec at the divertor while operating at elevant temperatures (800-1100 K or so) [1, 3-5]. With expected operational duty cycles the displacement damage from energetic neutron irradiation is anticipated be a few dpa/year which is expected to produce alloying by transmutants at a level of about 1 at. %/year and volumetric He production of ~5 appm/year in the wall and divertor armor materials[6].

Based on these considerations, tungsten-based alloys (W) have been proposed as a possible solid material for use in the first wall and divertor target regions[4, 7]. If the W thermal conductivity were to suffer degradation e.g. due to irradiation effects with a fixed target heat flux, it is likely that with existing designs[8, 9] near-surface recrystallization would occur, increasing the risk of brittle fracture and armor failure. Thus the evolution thermal conductivity under the combined action of plasma and displacement damage irradiation is important for credible divertor designs. Furthermore, the retention of T in the first wall and divertor is a well recognized additional challenge for both tritium inventory control and achieving a closed tritium fuel cycle[10, 11]. For a tritium breeding ratio TBR>1 and effective wall recycling coefficient R<1, a straightforward particle balance [12] shows that the probability of permanently trapping tritium in the material surfaces of the device, $p_{trap}$, must satisfy

$$p_{trap} < (TBR-1)(1-R)\frac{p_{burn}\eta_{fuel}}{1-p_{burn}\eta_{fuel}} \quad (7)$$

where $p_{burn}$<1 is the tritium burn-up probability and $\eta_{fuel}$<1 is the efficiency of fueling the core plasma region by injecting fuel across the plasma boundary (*n.b.* for simplicity this efficiency is assumed to hold for both freshly injected and recycled fuel atoms; accounting for the lower fueling efficiency of recycled atoms would lower the maximum allowable value for $p_{trap}$). With typical values TBR~1.05, R~0.99-0.999, $p_{burn}$~0.05, and $\eta_{fuel}$~0.2-0.3, we then estimate an upper limit of $p_{trap} < 10^{-6} - 10^{-7}$. Thus fusion fuel retention in radiation damaged W operating at reactor-relevant temperatures *during* displacement damage is of interest, since these temperatures may be high enough for annealing effects to partially mitigate damage effects. Motivated by these considerations, we present experiments focused on retention and thermo-mechanical property evolution in ion-beam damaged W materials.

## 2.0 Previous Work

Hydrogen isotope retention physics mechanisms in plasma-exposed tungsten have been extensively studied[10, 13]. Hydrogen isotopes are mobile in tungsten at the expected temperatures in fusion experiments[14]. Thus deuterium atoms implanted by the plasma in the near surface (few nm's) region can diffuse into the bulk region, forming a spatially decaying profile deeper into the material. As they diffuse, deuterium atoms can become trapped at grain boundary, dislocation, and vacancy defect sites as well as at precipitate (i.e. void, bubble or blister) sites. The de-trapping energy (i.e. the energy which a trapped particle must acquire to escape the trap) can vary considerably depending on both the type of trap as well as on the number of atoms already contained within the trap;

theoretical calculations suggest that it can vary from ~0.2-0.3eV up to 1.5eV or more[15-17] depending on the trap type.

The creation of new trap sites via energetic neutron bombardment and subsequent creation of displacement cascades will increase the density of trap sites and can thereby impact the retention of fusion fuel in the wall and divertor materials. Increased retention of plasma-implanted D was reported in light (D, He) ion-beam damaged single crystal [18] and polycrystalline [19] W. The results showed evidence of higher energy traps presumably from displacement damage-induced vacancies. The retention increase appeared to saturate when the damage level approached 0.4 dpa [20], and this increased retention observed at lower temperature was largely/wholly eliminated when damaged material was subsequently annealed at ~1200 K. Retention of D in self-damaged W (i.e. in W exposed to a W ion beam) due to subsequent high-flux D plasma exposures has been reported [21]. A marked increase in retention was reported for 300-500K plasma exposures, and the increase was found to saturate as the damage level approached 0.5 dpa. A much smaller increase in retention was reported when the sample temperature increased to 700-1000 K during damage. Similar results were reported in [22]; in this work *in-situ* TEM imaging also showed that smaller sizes defects merged into one another at elevated temperatures, forming a lower density of larger vacancy clusters. Similar results were reported when the plasma ion flux approaches values expected in divertor target of engineering test reactors[23]. Finally, the presence of damage also affected the observed D profile evolution[24], particularly with higher implanted D ion fluence, indicating that the transport of D within damaged material is affected by the presence of a new population of trap sites induced by the damage process.

Shimada and collaborators carried out plasma-implanted D retention studies in neutron-damaged W [25-30]. NRA was used to find the near-surface retained D and was compared to TDS. The TDS release data showed significantly more retained D than did NRA, providing indirect evidence for deep (>5 micron) D retention in neutron damaged W. These workers also reported a broader range of release temperatures of D during TDS, suggesting that the trap energy distribution is different in the two types of damage studies.

The literature on thermal diffusivity evolution due to displacement damage is much less extensive. Initial studies of thermal conductivity in He-ion beam displacement-damaged plasma facing tungsten armour [31] report significant (~factor of 2 or more) reduction in thermal diffusivity (and thus in thermal conductivity) for 0.2 dpa tungsten damaged by a He ion beam. The results were compared favorably with published models of defect scattering of electrons, the main heat carriers in W.

# 3.0 Experimental Techniques

For the retention experiments, W samples were prepared as documented[32], and were then damaged at the Ion Beam Materials Laboratory (IBML) at LANL. For the work here, a Cu ion beam with energy between 0.5-5.0 MeV was used to induce uniform displacement damage in the first ~1 μm of a W surface that was held at a controlled temperature ranging from 300-1200 K during ion beam exposure. The specific damage level corresponding to the ion beam fluence on the surface was determined from SRIM (Stopping and Range of Ions in Matter) Monte Carlo code calculations as described elsewhere [32]. Samples were then held at a temperature of 380K and exposed to a D plasma in a helicon RF-produced plasma device [33] (in these experiments the source was operated as an unmagnetized inductively coupled source). A 100 eV ion current density of $4x10^{20}$ ions/m$^2$-sec provided a fluence of $10^{24}$ ions/m$^2$. NRA profiles of trapped D were obtained using a $^3He^+$ ion beam with energies of 0.6, 0.8, 1.2, 1.6, 2.0, 2.5, and 3.5 MeV to obtain the optimum depth resolution at different depths. The results then provide D profiles with resolution of less than 1 μm up to a depth of ~6 μm in W. Thermal desorption spectroscopy (TDS) with a ramp-up of 0.5 K/sec was used to measure D release temperatures and total D inventory. A detailed discussion of these techniques is available [12],[32]. This particular choice of ion beam and plasma exposure protocol shows how simultaneous displacement damage and annealing at elevated temperatures affects the progressive release of trapped D, and thus provides some initial glimpse into how annealing affects lower energy traps present in undamaged W verses higher energy traps thought to be associated with displacement damage. We note that results obtained with a similar protocol have been reported [22, 34].

The thermal conductivity of Cu ion beam damaged W was studied using the $3\omega$ technique as described in the literature [35]. Briefly, a thin insulation layer of 30 nm thick Al2O3 was deposited via atomic layer deposition (ALD) at 520 K on the sample surface. A metal strip, made of 25 nm thick Cr and 125 nm thick Au, was deposited and patterned on top of the insulation layer. The metal strip works both as a heater and a thermometer. By varying the oscillation frequency the penetration depth of the transient heating can be varied, allowing the inference of thermal conductivity within the damage region located in the first micron of the surface.

# 4.0 Experimental Results

Figure 1 shows D retention results from TDS carried out with a ramp-rate of 0.5 K/sec for samples damaged at 300 K. The results show a clear increase in retention as the amount of damage is increased. This increase is particularly pronounced for the higher temperature release peaks, which are associated presumably with higher energy traps, as shown e.g. by the elevated release rate when the sample temperature reaches 700-1000K.

The relative increase of retention over that found in an undamaged sample obtained from both TDS and NRA measurements[32] shows retention proportional to $dpa^{0.66}$. These results confirm the results summarized in section 2 above, and provide a baseline for comparison with the following results.

The temperature of the sample during the displacement damage process play an important role in the total retention, as shown in Figure 2 for samples damaged to a level of 0.2 dpa. For a sample damaged at 300 K (blue curve, Figure 2) we observe two broad release features. The first occurs between 400-600 K, and a second release feature occurs between 700-1000K. This latter peak contains most of the retained inventory with a peak release rate occurring at about 850 K, consistent with the results in Figure 1. This higher temperature release peak is presumably associated with D that is found within more deeply trapped sites that were generated via the ion beam displacement damage process. Samples that were damaged at 1240 K show a retained D release profile that is nearly identical to the control TDS release profile, obtained for W samples that were implanted with the same plasma D ion fluence but which were not subject to any displacement damage. In particular, the high temperature release peak associated with the damage process has basically disappeared when the damage occurs at 1240 K, presumably via an annealing process that is activated at sufficiently high temperature. Our ion beam experiments have damage rates of order $10^{-4}$ dpa/sec which is several orders of magnitude higher than the rate of damage expected from neutron irradiation (which is estimated to be in the range of $10^{-7}$ dpa/sec). Thus it stands to reason that the rate of annealing would be more than enough to materially affect the density of neutron-damage induced vacancies. We can therefore conclude that if the PFCs are operated at sufficiently high temperature in a long pulse burning plasma device a similar annealing effect of the displacement-damage induced trap sites should be expected to occur at the lower damage rates found during neutron damage.

Measurements of the thermal conductivity of the damaged layer were also performed for Cu-ion beam damaged W samples. Here the damage profile was held uniform up to a depth of 1 micron by using multiple Cu ion beam energies ranging from 0.5-5.0 MeV to induce the desired dpa profile [32]. Results shown in Figure 3) indicate that samples damaged at room temperature exhibit a measurable decrease in thermal conductivity even at $10^{-2}$ dpa. The decrease reaches a value of roughly 50 W/m-K as the damage level reaches a value of 0.6 dpa, a factor of ~3x lower than the undamaged value.

## 5.0 Discussion

The retention results reported here are consistent with a number of previously reported findings. In particular, in samples damaged at low temperature, we have observed a clear increase in retention with increasing damage. This increased retention occurs in the region where displacement damage is computed to occur, and is associated with traps that release the D atoms at high (~850 K) temperature during TDS. We also find a clear decrease in the retention in these deeper traps when damage occurs at sufficiently high (~1000-1200 K) temperatures. These retention effects are observed at damage levels (~0.1dpa) that would correspond to only a few weeks of operation under projected DEMO device operations[6] and are large enough to impact tritium breeding and inventory control.

The earlier discussion of TBR motivates us to estimate the trapping probability, $p_{trap}$, which was introduced earlier in this paper as an important parameter that affects the achievable TBR. Recent work [36] provides measurements of retained D inventory vs plasma fluence for undamaged tungsten. Using these data, and taking $p_{trap}$ to be given then by the ratio of the retained D inventory to the total ion fluence to the surface, we can then examine the evolution of $p_{trap}$, vs. plasma ion fluence. Figure 4 shows this result for several types of plasma-exposed W. The open red circles show $p_{trap}$, vs. plasma ion fluence for undamaged W exposed to D plasmas in PISCES plasmas, where the W samples were held at 643 K. A detailed description of these PISCES exposures is available elsewhere [36]. These data show a precipitous drop in $p_{trap}$ with increased plasma fluence, presumably due to the fact that as the fluence is raised, the near-surface intrinsic traps become saturated with D. The resulting saturated near-surface region is then more likely to release D back to the plasma, thereby reducing $p_{trap}$. A fit to these data is also shown as the red line in Figure 4. As discussed elsewhere [36], the retention probability decreases significantly as the material temperature during plasma exposure is increased. Thus the retention probably vs fluence curve is expected to drop to much smaller levels e.g. as shown by the green line for W exposed to pure D plasmas at 1000 K, based on retention modeling [36].

The question then arises: how significant is the increased retention that occurs in moderately (<1 dpa) damaged W? The results presented in here permit us to begin to address this question. The data reported in this paper were obtained at a low plasma ion fluence of $10^{24}$ ions/m$^2$ across a range of W material temperatures during the damage process. When recast as an experimentally measured retention probability as defined above, the results from 0.2 dpa damaged W at room temperature shows a retention probability that greatly exceeds the maximum permissible $p_{trap}$ for TBR>1 as indicated by the blue open circle data point (for 380 K exposure temperature) and the filled blue circle data point (for 1200 K exposure temperature) in Figure 4. These data points lie well above the measured (red line) and modeled (green line) that correspond to the retention fraction for undamaged W exposed at 643 K and 1000 K to pure D plasmas. Thus our results suggest that damage levels of 0.2-0.3 dpa are sufficient to materially impact retention.

Published data from neutron irradiated samples [28] are also used to generate another radiation-damaged data point shown in Figure 4. This data point was obtained for a neutron exposure temperature of 320 K, nearly the same as our 380 K exposure data and low enough that annealing of displacement damage is negligible. These neutron exposed samples were then subjected to a plasma fluence of $10^{26}$ ions/cm$^2$, a value that is comparable to the PISCES-B data points shown in Figure 4 and the samples were then outgassed using TDS. The neutron exposed sample has a value of $p_{trap}$ that is about one order of magnitude higher than that of the PISCES data, and lies well above the TBR>1 ceiling for $p_{trap}$. The blue dashed line drawn between our lower fluence radiation damage result and the result from these higher ion-fluence neutron-irradiated samples gives an initial indication of how $p_{trap}$ will vary with ion fluence for moderately damaged (0.2-0.3 dpa) PFCs operated at low (320-380 K) temperatures These results show that the value of $p_{trap}$, in undamaged material appears to be small enough to allow TBR>1 once the fluence exceeds ~$10^{26}$ ions/m$^2$-sec which, assuming a first-wall ion flux of $10^{20}$ ions/m$^2$-sec, would occur in about two weeks of operation in DEMO. Further experiments in damaged W at higher temperature are necessary to determine the trapping probability at the higher first wall temperatures that would be anticipated to occur in a DEMO device.

Paranthetically, we note that as has been reported earlier [37], if the plasma has a mixture of D and He ions incident on the W surface, a thin (20-30 nanometer) layer of He-filled nanobubbles has been observed to form in undamaged W when the surface temperature is below a critical value. This nanobubble layer has been found to act as a strong diffusion barrier for plasma implanted D ions[37] and, as a result, the trapping probability drops to values well below the threshold needed for TBR>1, as shown by the black solid circle data points in Figure 4. At this writing it is unknown if a similar process will occur in displacement-damaged W.

As discussed in the introduction, we estimate that in order to maintain TBR>1, future devices will need to have $p_{trap}<10^{-6}$. These PISCES data show that for plasma fluences exceeding $10^{26}$ ions/cm$^2$, undamaged W exposed to pure D plasmas would appear to have sufficiently low retention to meet this criteria. Moderate amounts of damage (<1 dpa) can lead to a significant increase in $p_{trap.}$

The reduction in thermal conductivity reported here is quite significant, and is in reasonable agreement with recent results obtained in 0.2 dpa He-ion beam damaged W using a different non-contact technique to measure thermal diffusivity[31]. We note that the one published study of thermal conductivity in neutron-damaged W show only much more modest (~20-30%) reductions in thermal conductivity[38]. As discussed earlier, the performance of the divertor target is extremely sensitive to this parameter. Reductions in thermal conductivity might force a reduction in the maximum allowable heat flux to the target in order to avoid recrystallization of the near-surface region of the divertor target.

# 6.0 Conclusions

We have observed a clear increase in D retention in room temperature damaged tungsten up to levels approaching ~0.2 dpa. This increased retention occurs in the region where displacement damage is computed to occur, and is associated with traps that release the D atoms at high (~850 K) temperature. We also find a clear decrease in the retention in these deeper traps when damage occurs at sufficiently high (~1000-1200 K) temperatures. These retention effects are observed at damage levels (~0.2-0.3dpa) that would correspond to only a few weeks of operation under projected DEMO device operations and are large enough to be of significance for tritium breeding and inventory control. Further work is needed to determine the retention probability during higher temperature damage and higher ion fluences, and in particular, the impact of simultaneous D+He plasma ions on fuel retention in radiation damaged material is of interest and should be the focus of future experiments. We also observe a pronounced reduction in thermal conductivity in displacement damaged W that, should it occur in neutron damaged material, would have significant implications for the performance of a divertor target.

## Funding Sources:

This work supported by grants from the U.S. Department of Energy Office of Science DE-FG02-07ER54912, the University of California Office of the President 12-LR237801. This work was performed, in part, at the Center for Integrated Nanotechnologies, an Office of Science User Facility operated for the U.S. Department of Energy (DOE) Office of Science. Los Alamos National Laboratory is an affirmative action equal opportunity employer, and is operated by Los Alamos National Security, LLC, for the National Nuclear Security Administration of the U.S. Department of Energy under contract DE-AC52-06NA25396.Mas. Funding sources had no role in the experimental design, execution and analysis of these results.

## Figures

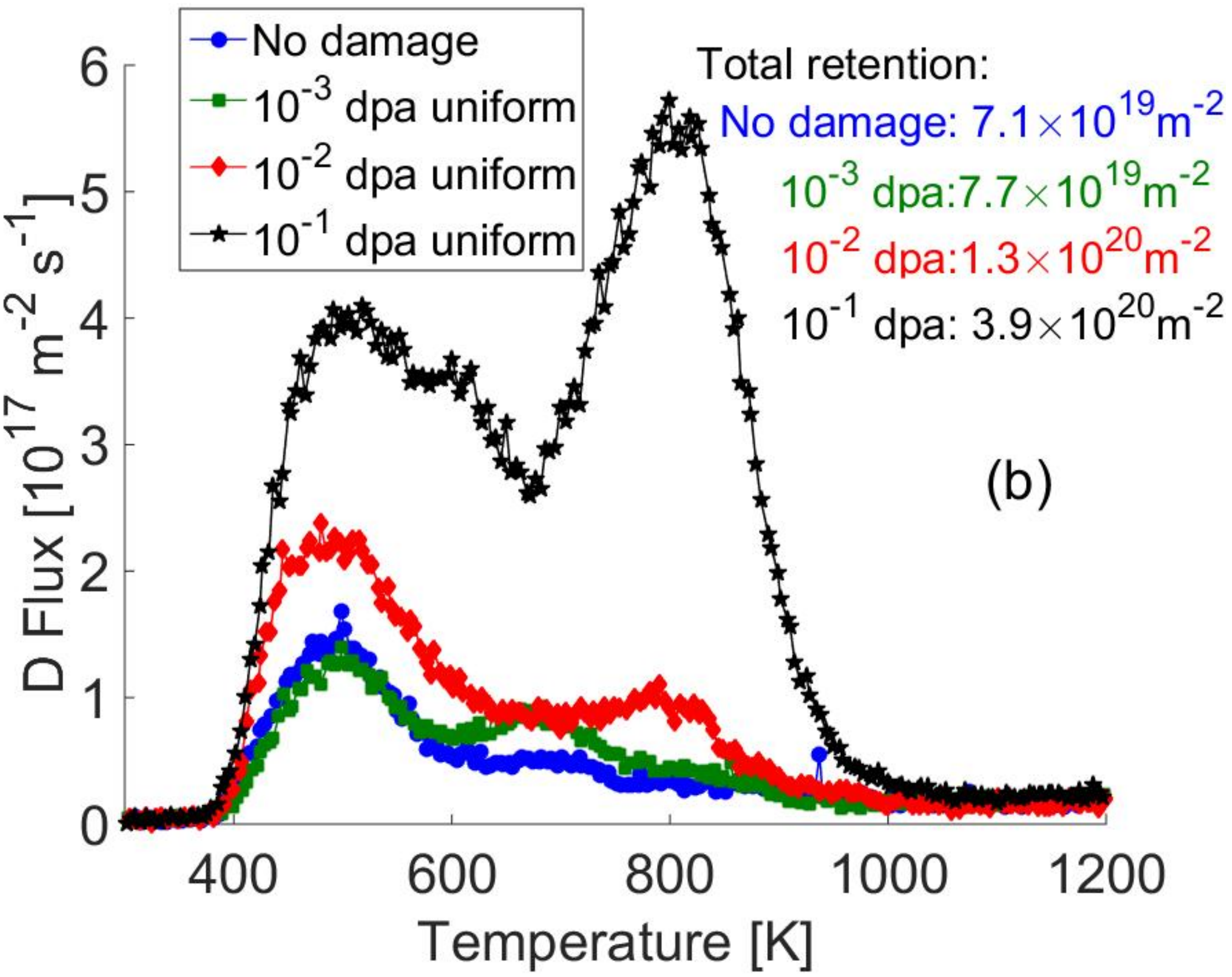

Figure 1: TDS D release rate vs temperature for W samples uniformly damaged by Cu ion beams to a depth of 1 micron while being held at a temperature of 300 K. Damaged samples were implanted with 100eV D plasma ions $10^{24}$ ions/m$^2$ fluence. Data obtained with a TDS temperature ramp rate of 0.5 K/sec. Pronounced increase in D retention is observed as the level of dpa increases, and a new D release peak at 800 K develops as the damage is increased.

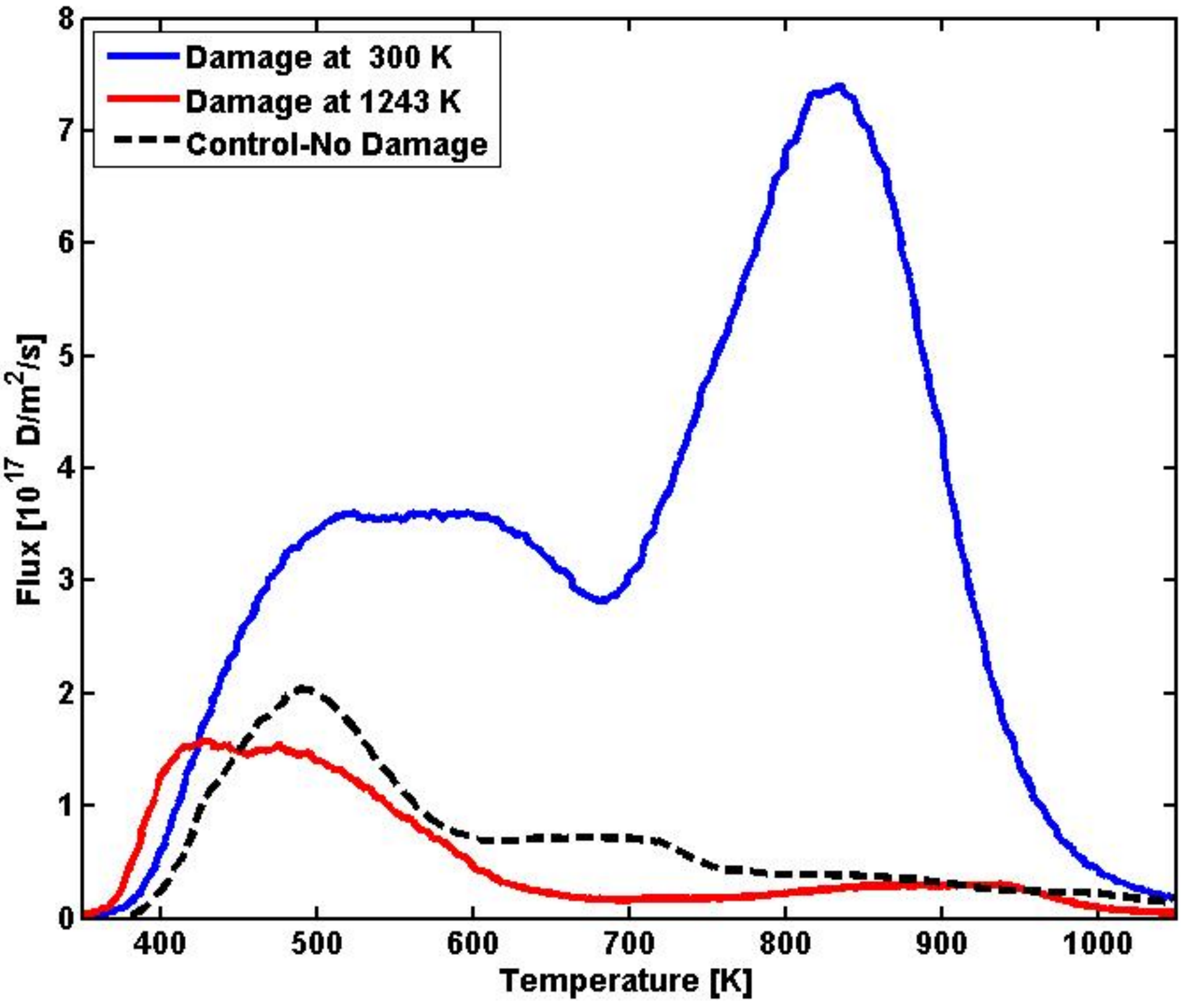

Figure 2: TDS emission spectra for undamaged sample (black dashed line), damaged to 0.2 dpa at 300 K (blue line) and at 1240 K (red line). Damaged samples were implanted with 100eV D plasma ions $10^{24}$ ions/m$^2$ fluence. Data obtained with a TDS temperature ramp rate of 0.5 K/sec. Retention is strongly reduced as sample damaging temperature is increased, with the largest reduction occurring in the higher release temperature peaks, indicating annealing rate is high enough to overcome ion beam damage rate.

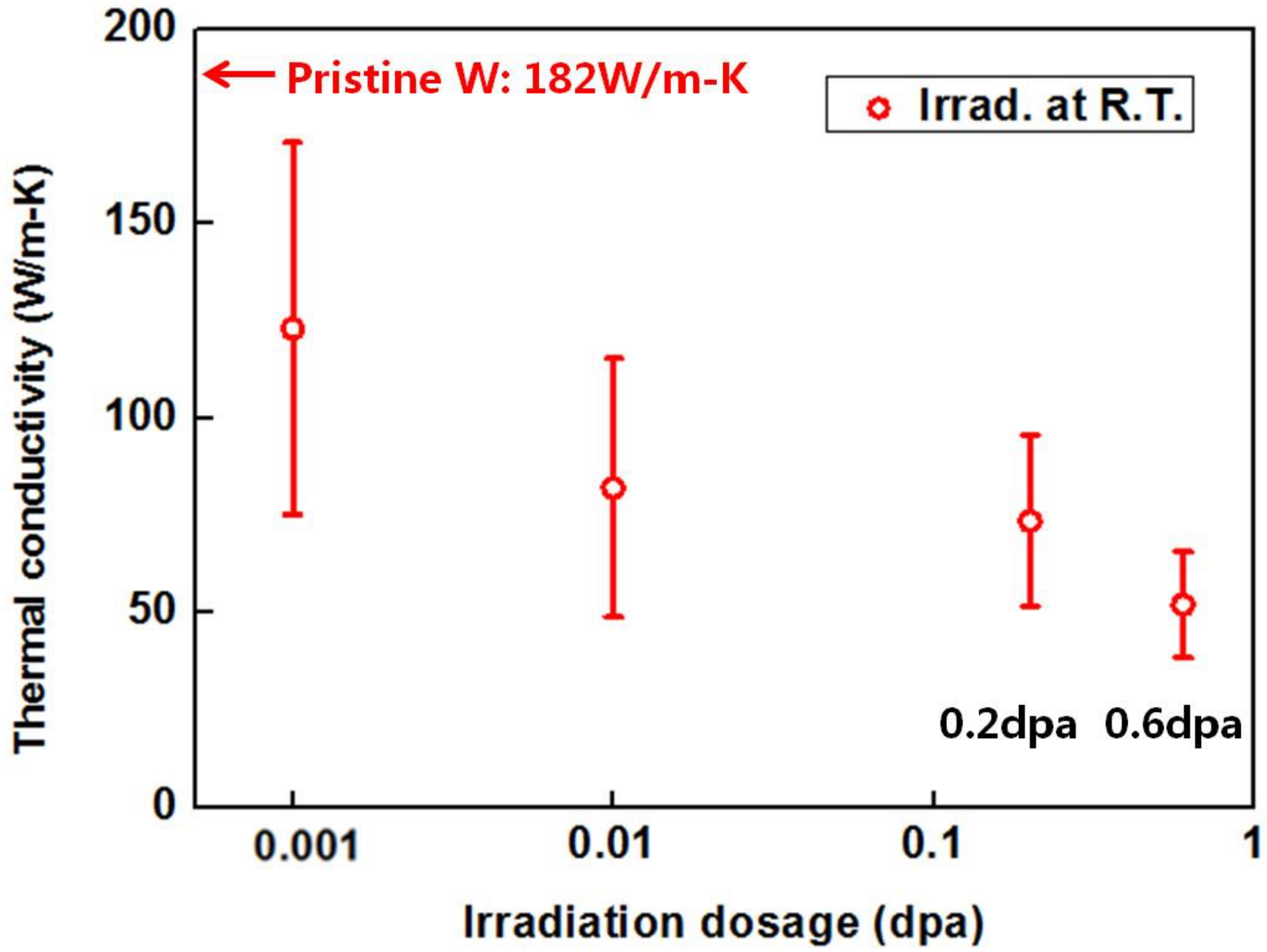

Figure 3: Thermal conductivity of damaged zone in W as measured by 3ω technique. A uniform damage profile of 1 micron depth was produced by multiple Cu ion beam energies . A pronounced decrease in the thermal conductivity occurs as the displacement damage increases.

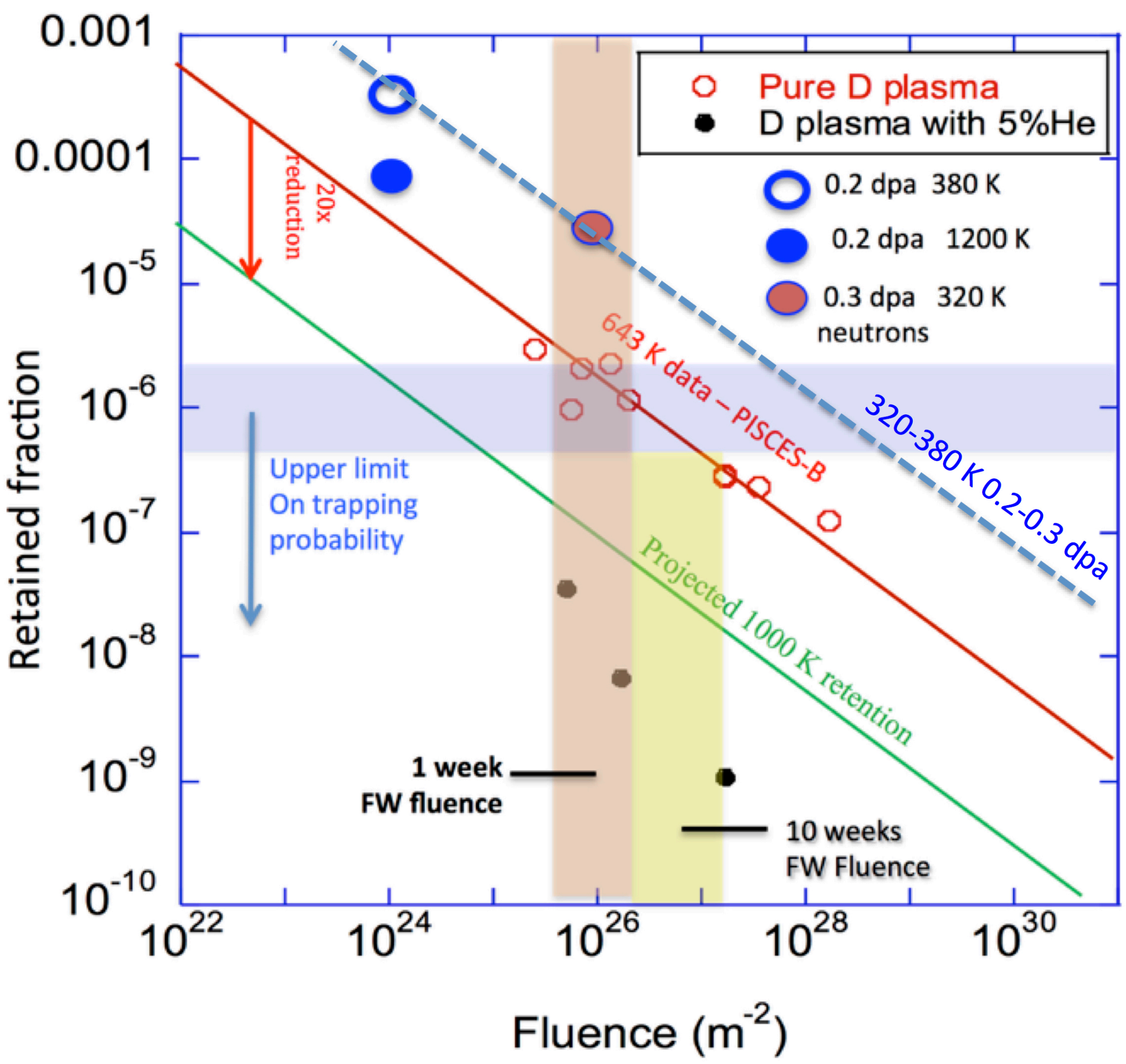

Figure 4: Retained fraction of deuterium in tungsten verses total plasma fluence. The retained fraction decreases as the total plasma fluence is increased, due to the gradual saturation of trap sites within the material. Existing experimental data in tungsten damaged to 0.2 dpa shown. Vertical colored areas show the expected fluence to a DEMO first wall for several operational periods for an assumed first-wall ion flux of $10^{20}$ ions/$m^2$-sec. Undamaged PISCES-B data points derived from published results[36]. Neutron data from Shimada et al[28].